\documentclass{iau} 
\usepackage{graphicx}
\usepackage{natbib}
\usepackage{hyperref}

\title[Thermonuclear X-ray bursts] 
{High-energy transients: thermonuclear (type-I) X-ray bursts}

\author[D. K. Galloway et al.]   
{Duncan K. Galloway$^{1,2}$,
Zac Johnston$^{1,2}$, Adelle Goodwin$^{1,2}$
 \and Alexander Heger$^{1,2,3}$}

\affiliation{$^1$School of Physics \& Astronomy, Monash University, Clayton, VIC 3800, Australia \\
$^2$Monash Centre for Astrophysics (MoCA), Monash University, Clayton, VIC 3800, Australia 
\\$^3$Tsung-Dao Lee Institute, Shanghai 200240, China
\\[\affilskip] email: {\tt duncan.galloway@monash.edu}}

\pubyear{2018}
\volume{339}  
\jname{Southern Horizons in Time Domain Astronomy}
\editors{A.C. Editor, B.D. Editor \& C.E. Editor, eds.}
\begin{document}

\maketitle

\begin{abstract}
Many distinct classes of high-energy variability have been observed in
astrophysical sources, on a range of timescales. The widest range
(spanning microseconds--decades) is found in accreting, stellar-mass
compact objects, including neutron stars and black holes. Neutron stars
are of particular observational interest, as they exhibit surface effects
giving rise to phenomena (thermonuclear bursts and pulsations) not seen in
black holes.

Here we briefly review the present understanding of 
thermonuclear (type-I) X-ray
bursts. These events are powered by an
extensive chain of nuclear reactions, which are in many cases unique to
these environments. 
Thermonuclear bursts
have been exploited over the last few years as an avenue to
measure the neutron star mass and radius, although the contribution of systematic errors to these measurements remains contentious. 
We describe recent efforts to better match burst
models to observations, with a view to resolving some of the astrophysical uncertainties related to these events. These efforts have good prospects for providing complementary information
to nuclear experiments. 
%
\end{abstract}

\firstsection 
\section{Introduction}

The high-energy sky is fundamentally dynamic.
From the early 1970s, when  X-ray (and higher-energy) bands opened to observers,  variability has been found on timescales of decades through to milliseconds.
This entire range is spanned by X-ray binaries, which show accretion rate variations on timescales of decades (including transient outbursts); through to ms timescales, including pulsations and quasi-perodic oscillations.

Low-mass binaries hosting neutron stars are thought to accrete through gigayear timescales \cite[e.g.][]{prp02}, sufficient to reduce the magnetic field to a point where it is dynamically unimportant. These systems exhibit a unique type of variability on timescales of seconds to minutes --- thermonuclear (type-I) X-ray bursts. 
Thermonuclear bursts occur when accreted fuel undergoes unstable ignition, producing bright ($\sim10^{38}\ {\rm erg\,s^{-1}}$) X-ray flashes  \cite[see][for a recent review]{gal17b}. Bursts typically ignite and burn via the triple-$\alpha$ reaction, and if hydrogen is  present, burn also via the 
$\alpha p$- and \textsl{rp}-processes. Much work has focussed on the \textsl{rp}-process, which can produce heavy proton-rich nuclei in the burst ashes \cite[e.g.][]{schatz01}. Many of the individual reactions have rates that are poorly measured experimentally, and involve nuclei with uncertain masses.

Although accretion rate and fuel composition are  the primary determinants of the burst properties, the burst luminosity profiles also encode information about the neutron star mass and radius (via the gravitational redshift) as well as the individual nuclear reactions that power them \cite[e.g.][]{cyburt16,so17}.

Much effort over the past decade has been to measure mass and radius from burst spectra \cite[e.g.][]{ozel16}. 
There remain fundamental uncertainties (and disagreements) about what bursts to choose, and what assumptions to make about the spectral shape;  \cite[e.g.][]{slb13,poutanen14}. 
%
Such issues are symptomatic of some remaining deep uncertainties about the burst physics, which motivate
further research (both observational and numerical) to resolve, and to improve our ability to constrain the properties of the burst hosts.

Here we describe the prospects for resolving these uncertainties via detailed comparisons of observations and numerical models.

\section{Observations}

Our knowledge of the phenomenology of thermonuclear bursts has grown from extensive observations made by a series of X-ray missions.
Notable examples include {\it BeppoSAX}, a mission featuring the Wide-Field Camera (WFC) operating through the 1990s \cite[]{boella97a,jager97,zand04b}; the {\it Rossi X-ray Timing Explorer} ({\it RXTE}), with the Proportional Counter Array (PCA) providing high sensitivity \& fast timing capability, and operational between 1995 December through 2012 January \cite[]{xte96}; and the 
 hard X-ray and $\gamma$-ray observatory {\it INTEGRAL}, with the wide-field Joint European X-ray Monitor (JEM-X), operational from 2002 onwards \cite[]{integral03,lund03}.

Other, currently-active missions with capabilities suited to observations of bursts include
{\it Swift}\/ \cite[]{swift04} \& {\it MAXI} \cite[]{maxi09}, each with wide-field instruments ideal for detecting new transients, and rare events like superbursts;
{\it NUSTAR} \cite[]{nustar10}, with sensitivity to ``hard'' X-rays (up to 80~keV);
{\it Insight-HXMT} \cite[]{zhang14}, launched in 2015 June, featuring three instruments providing a broad (1--250~keV) bandpass with high sensitivity;
{\it ASTROSAT} \cite[]{astrosat14}, launched in  2015 September, featuring the Large-Area X-ray Proportional Counter (LAXPC) with comparable capabilities to the {\it RXTE} PCA; and finally
{\it NICER} \cite[]{nicer16}, deployed to the International Space Station in 2017 June, with an observational program focussing on X-ray pulsations and bursts.

The data accumulated to date have revealed a remarkable diversity of burst behaviour. Amongst the usual frequent, quasi-regular bursts (lasting up to a minute and separated by a few hours), the most intense events usually exhibit
photospheric radius-expansion. Such events  are thought to reach the (local) Eddington flux limit, so that additional energy input goes into expansion of the photosphere. These bursts serve (approximately) as a standard candle allowing the distance to the bursting source to be estimated \cite[e.g.][]{kuul03a}. 
``Intermediate duration'' bursts, lasting minutes 
(and with correspondingly longer recurrence times) than typical bursts, are observed in low-accretion rate systems and are attributed to burning of large pure-He reservoirs \cite[e.g.][]{falanga09}. 
Even longer events lasting hours are classified as ``superbursts'', and are likely powered by carbon produced as a by-product of the burning during more frequent bursts \cite[]{corn00,zand17c}. 
%
%
Multi-peaked bursts have been suggested to arise as a result of ``nuclear waiting points'', specific reactions through which the burning products flow particularly slowly \cite[e.g.][]{fisker04}. 

\section{Analysis of large burst samples}

Given the diversity of burst phenomenology, assembly and analysis of large samples of bursts are important for identifying suitable candidates for analysis \cite[e.g.][]{corn03a,bcatalog}.
A more recent project, 
the Multi-INstrument Burst ARchive (MINBAR\footnote{\url{http://burst.sci.monash.edu/minbar}}), seeks to combine data from {\it multiple} instruments, and is currently under assembly.
The  MINBAR sample includes bursts observed by {\it BeppoSAX}/WFC, {\it RXTE}/PCA, and {\it INTEGRAL}/JEM-X, through to the end of the {\it RXTE}\/ mission in 2012 January. The sample will comprise more than 7000 events from 85 (of 110 known) sources. Analysis of the sample is expected to provide
an improved ``global'' view of burst behaviour and increased numbers of rare events. 


Analysis of preliminary sample data have already led to some significant results.
Spectral analysis of the bursts observed with the highest-sensitivity instruments suggests that the persistent emission increases temporarily during bursts \cite[]{worpel13a,worpel15}. This result has been 
corroborated with a burst fortuitously observed simultaneously with the {\it Chandra X-ray Observatory}\/ \cite[]{zand13a}, 
demonstrating that the increase factor can be as high as 20. Such an increase may be expected due to a temporary increase in accretion through the disk, resulting from
Poynting-Robertson drag on the disk material by the burst.
This result further suggests that the the traditional approach for time-resolved spectroscopy, requiring subtraction of the pre-burst emission, and fit with a blackbody \cite[e.g.][]{kuul02a}, may be inadequate for very high quality data.

Short recurrence-time bursts have been shown to be associated only with systems that accrete hydrogen-rich material, supporting the view that these events arise from fuel that is left unburnt from the previous event \cite[]{keek10}.
A survey of the properties of burst oscillations, transient quasi-periodic intensity variations detectable around the peak of bursts from some sources, has been presented in \cite{ootes17a}. 
%

\section{Burst models}

It is generally not possible to  infer the system parameters (neutron star mass, radius, fuel composition, etc.) directly from the bursts or the persistent emission observations. Consequently, we must make comparisons with burst models.
The current state-of-the-art is represented by 1-D codes  with adaptive nuclear reaction grids, like {\sc KEPLER} \cite[]{woos04} and {\sc MESA} \cite[]{mesa15}.

Much of the modelling effort to date has focussed on matching the behaviour of GS~1826$-$24, the ``Clocked Burster'', unique amongst burst sources for its consistent, regular bursts with uniform lightcurves. This behaviour has been observed over a range of accretion rates sampled at different epochs \cite[e.g.][]{clock99}.
Early comparisons of burst properties measured by {\it RXTE}\/ with simple ignition models suggesting low-metallicity fuel \cite[]{gal03d}. 
In contrast, subsequent analyses, also incorporating comparison between an observed lightcurve and a {\sc KEPLER} model result, indicted instead solar composition \cite[]{heger07b}. 
%
More recent, ongoing work with more comprehensive comparisons suggested that the degree of agreement may have been overestimated due to the restriction of comparison at a single epoch.

Efforts to improve the fidelity of these codes are ongoing, for example via model cross-comparisions coordinated via working groups assembled as part of the activities of the Joint Institute of Nuclear Astrophysics Centre for the Evolution of the Elements (JINA-CEE\footnote{\url{http://jinaweb.org}}).
There is also a need to improve the access to model results for observers. For example, a large sample of {\sc KEPLER} model results has been analysed by \citealt{lampe16}, with various parameters (burst recurrence time, lightcurves, etc.) made available via website\footnote{\url{http://burst.sci.monash.edu/kepler}}.

In order to anticipate the likely impact that model-observation comparisons can have on the rates of the nuclear reactions which drive the burst, sensitivity studies have been carried out to identify those reactions which have the most influence on the burst lightcurve \cite[]{cyburt16} 

\section{Recent work and astrophysical uncertainties}

Here we examine some of the recent results that demonstrate the potential of the model-observation comparisons, as well as highlighting some of the remaining issues that must be considered.
The ultimate goal is to match models to observations, taking into account the astrophysical uncertainties, and hence probe the nuclear physics of the bursts. 

Obtaining suitable data for model-observation comparisons can be a challenge. Low-Earth orbit satellite data is typically interrupted every $\approx90$~min satellite orbit, with maximal duty cycles of approximately 60\%. These data gaps introduce ambiguities for measuring recurrence times of typically a few hours. Systems other than GS~1826$-$24 typically exhibit much less regular bursts, and also may not span a sufficient range in accretion rate to provide burst samples with different ignition conditions.
To address this challenge, we have assembled from the MINBAR sample a set of observed bursts with well-constrained recurrence times \cite[][; see also http://burst.sci.monash.edu/reference]{gal17a}. It is anticipated that this sub-sample will serve as test cases for numerical codes to understand variations between models.
Where feasible, this sample includes observations at different accretion rates, specifically to enable multi-epoch comparisons to resolve astrophysical uncertainties.

\begin{figure}[t]
\begin{center}
 \includegraphics[width=3.4in]{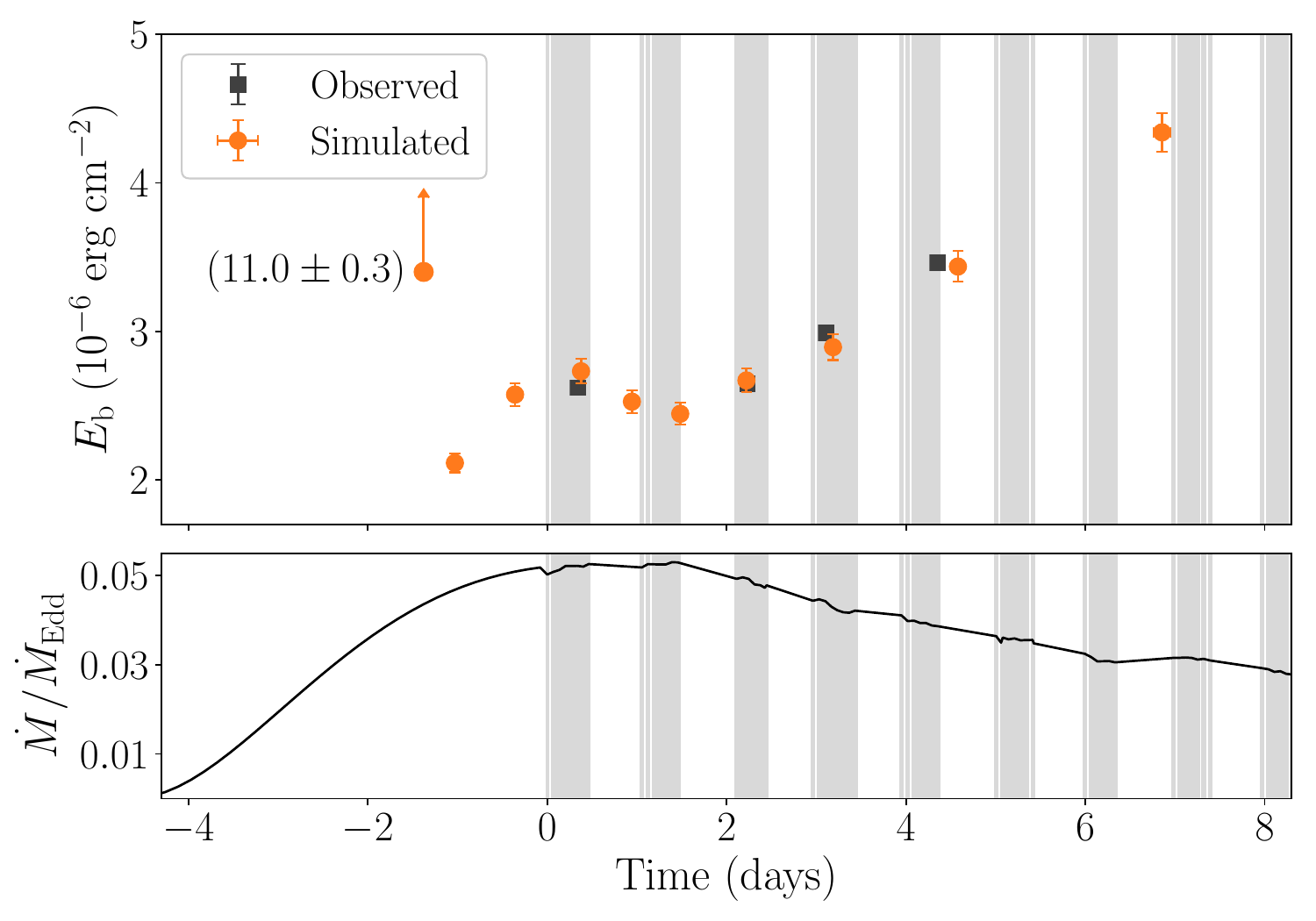} 
 \caption{Results from time-dependent {\sc KEPLER} model simulations of bursting activity during the 2002 outburst of SAX~J1808.4$-$3658. The upper panel plots the fluence, $E_b$, of a modelled burst sequence ({\it orange symbols}) for comparison against the four observed bursts ({\it black symbols}). Note the good agreement between the predicted fluence and time of the bursts; the vertical grey bands indicate the instrumental observation windows. The lower panel shows the inferred evolution of the accretion rate over the outburst, plotted as a fraction of the Eddington rate. From \cite{johnston17}}
   \label{fig1}
\end{center}
\end{figure}

An alternative target is short-duration transients, which may exhibit increases in their accretion rate by orders of magnitude over a day, and then decline again into quiescence. The accretion-powered millisecond pulsar SAX~J1808.4$-$3658 exhibits such outbursts every few years, and has previously been a target for comparisons of observed bursts with simple ignition models \cite[]{gal06c}.
More recently, the inferred mass accretion history during such outbursts has been used as input to {\sc KEPLER} to simulate the bursts observed over an entire week-long outburst \cite[][see also Figure \ref{fig1}]{johnston17}.
Such simulations demonstrate systematic differences in predicted recurrence times when the accretion rate is rising or falling, compared to the predicted recurrence time adopting the average accretion rate over the burst interval. This result may have implications for the response of the burning layer to varying accretion rate.

Such model-observation comparisons yet face significant challenges. Astrophysical uncertainties introduce biases and degeneracy in the comparisons. For example, the 
distance to the bursting sources are typically poorly known, introducing uncertainties to the burst energetics \cite[e.g.][]{bcatalog}. 
The measured burst flux is expected to be enhanced or attenuated 
due to the anisotropy of the environment  \cite[the accretion disk; e.g.][]{he16}. 
Estimates of the accretion rate are made via the persistent emission, which suffers the same problem (but with a different geometry factor). Additionally, the burst and persistent intensity is typically measured over a limited instrumental passband, introducing additional errors when estimating the bolometric luminosities \cite[e.g.][]{thompson08}.
The burst emission is also affected by gravitational redshift, which is poorly constrained due to uncertainties in mass and radius.

\section{Summary and future prospects}

There remain some fundamental shortcomings in our understanding of the various burst phenomena. We now have access to a substantial accumulated observational dataset to analyse, as well as multiple model codes with which to simulate bursts and hence infer system parameters. 
Software development is under way to provide code that can model these various effects and hence marginalise out the corresponding systematic uncertainties, to constrain the properties of interest.
As a result, the prospects for future model-observation comparisons are excellent, and incorporating known sensitivities to the nuclear physics may allow us to constrain specific masses and/or reaction rates, providing complementary information to nuclear experiments. Additionally, we have the continuing prospect of exciting new data obtained from new and upcoming missions incuding 
{\it Insight-HXMT},
{\it ASTROSAT} and {\it NICER}.

\acknowledgements

I would like to thank the organisers for their invitation and a superbly organised meeting on a timely topic.
The Multi-INstrument Burst ARchive (MINBAR) collaboration acknowledge the support of the Australian Academy of Science via its Scientific Visits to Europe program; the Australian Research Council's Discovery Projects and Future Fellowship funding schemes (DG \& AH); the US National Science Foundation under Grant No. PHY-1430152 (JINA Center for the Evolution of the Elements); the International Space Science Institute in Bern, Switzerland; and the European Union's Horizon 2020 Programme under the AHEAD project (grant agreement n. 654215).

\bibliography{all}
\bibliographystyle{apj}

%


\end{document}